\documentclass[12pt,a4paper]{article}
\usepackage{amsmath}
\usepackage{algorithm}
\usepackage{algpseudocode}
\usepackage{geometry}
\usepackage{graphicx}
\DeclareMathSymbol{\mg}{\mathrel}{symbols}{"1D}

\renewcommand{\Im}{\text{Im}\ }

\newcommand{\beq}{\begin{equation}}
\newcommand{\eeq}{\end{equation}}
\newcommand{\barr}{\begin{array}}
\newcommand{\earr}{\end{array}}

\newcounter{oldcounter}

\newcommand{\bgl}{{\bar\lambda}}

\newcommand{\bgps}{{\bar\psi}}

\newcommand{\Bga}{{\boldsymbol \alpha}}
\newcommand{\Bgb}{{\boldsymbol \beta}}

\usepackage[utf8]{inputenc}

\usepackage[T1]{fontenc}
\RequirePackage{amsmath,amssymb}
\RequirePackage{mathrsfs}
\RequirePackage{amsfonts,bm}
\RequirePackage{dsfont}
\RequirePackage{braket}
\RequirePackage{tabularx}
\usepackage{multirow}
\RequirePackage{nicefrac}
\RequirePackage{graphicx}
\RequirePackage{booktabs}
\RequirePackage{colortbl}
\RequirePackage{xcolor}
\RequirePackage[symbol]{footmisc}
\usepackage{mathtools}
\usepackage{comment}
\usepackage{blkarray}
\usepackage{float}
\usepackage{rotating}
\usepackage{hhline}
\usepackage{tikz}
\usepackage{multirow}
\usepackage[section]{placeins}

\usepackage[natbib=true,maxbibnames=99,giveninits=true,sorting=none,safeinputenc]{biblatex}

\DeclareUnicodeCharacter{2032}{'}

\def\beq{\begin{equation}}
\def\eeq{\end{equation}}
\def\beqn{\begin{eqnarray}}
\def\eeqn{\end{eqnarray}}

\usepackage{empheq}
\usepackage[most]{tcolorbox}
\usepackage{blkarray}
\newtcbox{\mymath}[1][]{%
    nobeforeafter, math upper, tcbox raise base,
    enhanced, colframe=blue!30!black,
    colback=blue!30, boxrule=1pt,
    #1}

\newcommand{\CC}[2]{C{#1\atopwithdelims[]#2}}

\newcommand{\ba}{\begin{eqnarray}}
\newcommand{\ea}{\end{eqnarray}}

\newcommand{\one}{\mathds{1}} 
\newcommand{\Sv}{\bm{S}}

\newcommand\St{\mathbf{\tilde{S}}}

\numberwithin{equation}{section}

\begin{document}
\begin{titlepage}
\samepage{
\setcounter{page}{1}
\rightline{July 2024}

\vfill
\begin{center}
  {\Large \bf{
     Vacuum Energy of \\ \medskip
        Non--Supersymmetric $\bm{\tilde{S}}$ Heterotic String Models 
 }}

\vspace{1cm}
\vfill

{\large Luke A. Detraux $^{1}$\footnote{E-mail address: ldetraux@liverpool.ac.uk}, \\ \medskip
Alonzo R. Diaz Avalos$^{1}$\footnote{E-mail address: a.diaz-avalos@liverpool.ac.uk}, 
Alon E. Faraggi$^{1}$\footnote{E-mail address: alon.faraggi@liverpool.ac.uk}
and
Benjamin Percival$^{1,2}$\footnote{E-mail address: b.percival@mmu.ac.uk}}

\vspace{1cm}

{\it $^{1}$ Dept.\ of Mathematical Sciences, University of Liverpool, Liverpool
L69 7ZL, UK\\}
\vspace{.08in}

{\it $^{2}$ Dept. of Natural Sciences, Manchester Metropolitan University,  M15 6BH, UK \\}

\vspace{.025in}
\end{center}

\vfill

\begin{abstract}

We use the free fermionic formulation of the heterotic--string in four dimensions
to study the vacuum structure and energy of non--supersymmetric tachyon free models that correspond 
to compactifications of tachyonic vacua of the ten dimensional heterotic--string. 
We explore the class of heterotic $SO(10)$ non--supersymmetric models constructed from the $\tilde{S}$-model in the Free Fermionic Formalism, and investigate the dependence of the potential on the geometric moduli.
This paper will explore a sample of $ 10^{9}$ string vacua to find the frequency of viable models, classifying these vacua by the following fertility criteria: tachyon presence; number of spinorial $\boldsymbol{16/\overline{16}}$ representations; vectorial $\boldsymbol{10}$ states; Top Quark Mass Coupling compatibility. Of these we find those that mimic supersymmetric models with equal number of bosons and fermions at the massless level - $a_{00} = 0$.  Tachyon free models occur with a frequency of $5.309\times10^{-3}$. Furthermore, models that fulfil the rest of the phenomenological fertility conditions and the additional condition on $a_{00}$ occur with probability $4.0 \times 10^{-9}$
We analyse the partition functions and study the moduli dependence of such models, finding that almost all fertile models have finite, positive potential at the Free Fermionic Point, with $2$ out of $84$ of the fertile cores having negative, finite potential. We demonstrate that the Free Fermionic Point is
not necessarily a minimum in the potential. 
This work provides further evidence that supersymmetry may not be a necessary ingredient of phenomenological models, recreating many of the desirable features of such models without employing supersymmetry. 
\end{abstract}
\smallskip}

\end{titlepage}
\newpage

\tableofcontents

\section{Introduction}
\label{sec:Intro}

The Standard Model (SM) of particle physics provides viable parameterisation of all observable 
data to date concerning particles and their interactions via three of the four fundamental forces, and it may remain viable up to the Grand Unified Theory scale, or the Planck scale, 
where the gravitational effects are non--negligible. However the SM is incomplete, and one of the outstanding problems in this construction is the Cosmological Constant Problem. String theory acts as a unique
laboratory to explore the synthesis of the gauge and gravitational interactions within
a self--consistent framework, and probe the open questions in the SM. Phenomenological string models are constructed
that reproduce the main phenomenological characteristics of the
Minimal Supersymmetry Standard Model (MSSM) \cite{fny,slm1,cfn}. The early
constructions since the late $1980$s generally possessed $N=1$ spacetime supersymmetry that guarantees 
the stability of the vacuum and simplifies the analysis in some respect by guaranteeing a vanishing Cosmological Constant at one-loop. 

More recently 
non--supersymmetric string vacua have been of interest as well
\cite{aafs, ADM, nonsusy5, CoCSuppression2, boyle2024classification, blaszczyk2014non}, and the stability of their potential have been investigated infrequently \cite{florakis2016chiral, avalos2023d, avalos2023fayet, angelantonj2024non}. 
These studies mainly focus
on compactifications of the $SO(16)\times SO(16)$ heterotic--string,
which is tachyon free in ten dimensions \cite{KLTClass, DH, AGMV, gv, nonsusy2}. In general, however, 
compactifications of the $SO(16)\times SO(16)$ heterotic--string to four dimensions give
rise to physical tachyonic states in the spectrum of the four dimensional models, which are 
projected from the spectrum by Generalised Gliozi--Scherk--Olive (GGSO) projections. In addition, 
heterotic string theory gives rise 
to non--supersymmetric vacua that are tachyonic in ten dimensions \cite{KLTClass, DH}. 
Physical tachyons can then be projected from the spectrum in the compactified theory 
by GGSO projections \cite{spwsp, stable} as in the $SO(16)\times SO(16)$ case, and so it is prudent to study the phenomenological 
properties and stability of these vacua as well.
Over the past few years several studies along these lines
has been pursued
\cite{spwsp, stable, faraggi2020towards,faraggi2021classification, fpsw}, 
as well as studies of non--supersymmetric four dimensional models that
correspond to compactifications of the $SO(16)\times SO(16)$ heterotic--string 
\cite{aafs,ADM,florakis2016chiral, CoCSuppression1,florakis2022super,VafaetalAsy1,VafaetalAsy2}. 
A specific class of these studies are on internal spaces that correspond to $\mathbb{Z}_2 \times \mathbb{Z}_2$
orbifolds of six dimensional toroidal lattices, and are analysed by using the Free Fermionic
Formulation (FFF) \cite{antoniadis19884d, KLT}  of the heterotic--string in four dimensions. Within these models, those that 
descend from the $SO(16)\times SO(16)$ heterotic--string are dubbed as 
$S$--models 
, whereas
those that descend from a specific
tachyonic ten dimensional vacuum are dubbed as 
${\tilde S}$--models.
We refer the readers to the literature for more detailed explanation of this nomenclature
\cite{spwsp, stable}. 
%
To date the value of the cosmological constant and its dependence on the structure of the vacuum has been 
exclusively studied in the context of 
S--models\footnote{We remark that S--models contain both non--supersymmetric
models, in which $N=1$ spacetime supersymmetry is broken by a GGSO phase, and
the supersymmetric models, in which the $\Sv$ basis vector is the spacetime 
supersymmetry generator \cite{spwsp}}. 
We extend the analysis to ${\tilde{S}}$--models. 

 In this paper we will build a non--supersymmetric $SO(10)$ models with the 
 $\bm{\tilde{S}}$ vector as one of 12 basis vectors based on the 
 $\overline{\rm NAHE}$ set using the Free Fermionic Formalism. Details of the construction are given in Section \ref{sec:S tilde}. From the possible $2^{66}$ independent GGSO phase matrices, our program, written in C++, takes a random sample of $ 10^{9}$ matrices and classifies these models by the fertility conditions developed in \cite{faraggi2020towards}. These fertility conditions are: level-matched tachyon projection; compatibility with the three generations of the SM fermions; compatibility with the Higgs doublet; maintaining Top Quark Mass Coupling (TQMC). With comparison to supersymmetric models in mind, we also identify models with massless fermion and boson cancellation, $a_{00} = N_{B}-N_{F} = 0$. Section \ref{sec: tachyons} details the origins of the problematic tachyonic sectors and their survival conditions, followed by analysis of the massless spectrum in Section \ref{sec: Massless}. 
 
 In the second half of this paper we move away from the Free Fermionic Point, (FFP), referred to as the fermionic point or fermionic radius in \cite{FRtranslation}, to analyse the partition function and resultant potential of these fertile cores. 
 This is achieved by bosonising the internal worldsheet fermions,  through the relationship
 $\partial X^{I} \approx \sqrt{\alpha'}y^{I}\omega^{I}$, where the fermionic radius $R = \sqrt{\frac{\alpha'}{2}}$. 
 We demonstrate that stability of a model is not exclusive to $S$--models and many of the features of supersymmetric and non--supersymmetric  
 $S$--models can be replicated in non--supersymmetric 
 ${\tilde{S}}$--models. 
 In particular, we demonstrate that stable minima in the potential can be found for non--supersymmetric $\tilde{S}$--models at or close to the FFP, when the geometric moduli, $T^{(1)}$ and $U^{(1)}$ are varied in imaginary direction. 
 Furthermore, we demonstrate that supersymmetry is not a necessary requirement for a vanishing $a_{00} = N_{f} - N_{b}$, with models obeying this constraint appearing sporadically in the sample of models we analyse, and giving finite, non-vanishing cosmological constant. In Sections \ref{sec:Partfunc} and \ref{sec:Pot}, we give an overview of how we constructed and analysed the partition function and potential of the fertile models at, and away from, the FFP, and how this relates to the Cosmological Constant. Finally, in Section \ref{sec:Results} we present the statistics that result from the classification and in Section \ref{sec:Ex} we give an example of a model that fulfils all fertility criteria and sits at a minima in potential close to, but not at, the FFP. Section \ref{sec:con} concludes this work with discussion of the results and prospective future work.

\section{The $\bm{\tilde{S}}$--Models}
\label{sec:S tilde}
The FFF builds a models spectrum up from a set of basis vectors. These basis vectors enforce boundary conditions on a set of world sheet fermionic degrees of freedom, used to cancel the conformal anomaly. The second ingredient in this formalism are the phases between basis vectors, given by a generalised GSO (GGSO) matrix. This method has been used to build toy models, corresponding to $\mathbb{Z}_2 \times \mathbb{Z}_2$ orbifolds of $6$ dimensional tori, for decades and has led to such discoveries as spinor-vector duality\cite{faraggi2007spinor}. Here we employ it to sample the space of non--supersymmetric vacua and classify the solutions phenomenologically. 
Firstly we will define our basis vectors, determining the configuration of the world sheet fermions in each sector. The following vectors are 
adapted from the $SO(10)$ classification basis \cite{fknr} by implementing the 
$\Sv\rightarrow \bm{\tilde{S}}$ map \cite{faraggi2020towards} and define our $\tilde{S}$--models in 
four dimensions. 

\begin{equation}
\begin{split}
\label{eq:BV}
&\mathds{1} = \mathbf{v}_{1} =  \{\psi^{\mu}, \chi^{1,...,6}, 
y^{1,...,6}, w^{1,...,6} ~| ~ \bar y^{1,...,6},\bar w^{1,...,6}~,~
\bar \eta^{1,2,3}, \bar \psi^{1,...,5}, \bar \phi^{1,...,8} \} \\
&\bm{\tilde{S}} = \mathbf{v}_{2} = \{\psi^{\mu}, \chi^{1,2}, \chi^{3,4},\chi^{5,6} | \bar \phi^{3,4,5,6} \} \\
&\mathbf{e}_{i} = \mathbf{v}_{2+i} =\{y^{i}, w^{i} | \bar y^{i},\bar w^{i}\} ~~~~~~~~~~~~~~~ \text{ for } ~~~~~~i = 1, 2, 3, 4, 5, 6 \\
&\mathbf{b}_{1} = \mathbf{v}_{9} =   \{\psi^{\mu}, \chi^{1,2}, y^{3,4,5,6} | \bar y^{3,4,5,6},\bar \eta^{1}, \bar \psi^{1,2,3,4,5} \} \\
&\mathbf{b}_{2} = \mathbf{v}_{10} = \{\psi^{\mu}, \chi^{3,4}, y^{1,2}, w^{5,6} | \bar y^{1,2},\bar w^{5,6},\bar \eta^{2}, \bar \psi^{1,2,3,4,5} \} \\
&\mathbf{z}_{1} = \mathbf{v}_{11} =\{\bar \phi^{1,2,3,4} \} \\
&\mathbf{z}_{2} = \mathbf{v}_{12} = \{\bar \phi^{5,6,7,8} \} \\
\end{split}
\end{equation}

Here, $\psi^{\mu}$ are the fermionic superpartners of the left--moving spacetime
bosonic coordinates in the light--cone gauge; 
$\chi^{1,...,6}$ are the fermionic superpartners of the $6$ dimensional left-moving 
compactified coordinates, $y^{i},w^{i}$; $\bar{y}^{i},\bar{w}^{i}$ correspond to the $6$ 
right--moving internal, fermionised coordinates; and $\bar{\eta}^{1,2,3}$, $\bar{\psi}^{1,2,3,4,5}$, and $\bar{\phi}^{1,2,3,4,5,6,7,8}$ are $16$ right-moving complex fermions required to cancel the conformal anomaly. 

Considering the basis vectors: 
$ \mathds{1}$ is required for consistency;
$\mathbf{\mathbf{e}_{i}}$ relate to shifts in the six dimensional compactified torus
\cite{fknr, Viktorthesis, FRtranslation}; 
$\mathbf{b_{a}}$ relate to $\mathbb{Z}_{2} \times \mathbb{Z}_{2}$ orbifold twists; and $\mathbf{\mathbf{z}_{a}}$ relate to hidden sector gauge group breaking and enhancements, where $a=1,2$.

However the defining feature of this model is the modification of the spacetime supersymmetry generating vector $ \Sv \rightarrow \bm{\tilde{S}}$,
\begin{equation}
    \label{eq:S} \Sv = \{\psi^{\mu}, \chi^{1,2}, \chi^{3,4},\chi^{5,6}  \} \rightarrow \bm{\tilde{S}} = \{\psi^{\mu}, \chi^{1,2}, \chi^{3,4},\chi^{5,6} | \bar{\phi}^{3,4,5,6}\}
\end{equation}

\noindent 
with the $\bm{\tilde{S}}$ vector removing supersymmetry from the model at the ten dimensional level, eliminating the massless gravitinos found in similar $S$-models. This mapping eliminates the need for a spontaneous supersymmetry breaking mechanism, and alters the structure of the model significantly.

Having defined our basis vectors, the features of the model now depend only on the GGSO matrix with components C$\begin{bmatrix}
\mathbf{v}_{i} \\
\mathbf{v}_{j}  \\
\end{bmatrix} = \pm 1 $, from which we construct the Hilbert Space
of the model,
\begin{equation}
    \label{eq:Hil}
\mathcal{H} = \oplus \prod^{k}_{i=1} \{ e^{i \pi v_{i} F_{\alpha} } | S_{\alpha} \rangle =\delta_{\alpha} C\begin{bmatrix}
\mathbf{v}_{i} \\
\mathbf{v}_{j}  \\

\end{bmatrix}^{*} | S_{\alpha} \rangle \}.
\end{equation}
where $v_{i}$ are the basis vectors, $\alpha$ labels the sector, $ F_{\alpha}$ is the fermion number operator, and $\delta_{\alpha}$ is the spin statistics index, equal to $\pm 1$.

Sectors, $|S_{\alpha} \rangle $, are constructed from linear combinations of the basis vectors (addition modulo 2) and are described by their mass. The left and right moving components of the mass are:
\begin{equation}
    \label{eq:ML}
    \mathcal{M}^{2}_{L} = -\frac{1}{2} + \frac{\alpha_{L} \cdot \alpha_{L}}{8}  +N_{L} 
    \end{equation}
\begin{equation}
    \label{eq:MR}
    \mathcal{M}^{2}_{R} = -1 + \frac{\alpha_{R} \cdot \alpha_{R}}{8}  + N_{R} \text{ ,}
    \end{equation}

\noindent where $N_{L}$, $N_{R}$ sum over the left-- and right--moving oscillators. Physical states satisfy the Virasoro level-matching condition $\mathcal{M}^{2}_{L} = \mathcal{M}^{2}_{R}$ 

As the entries above the diagonal of the $12 \times 12 $ GGSO matrix are free parameters, taking values of $\pm 1$, a space of $2^{66}$ potential GGSO matrices describe $2^{66} \approx 10^{20}$ potential models. Only a fraction of these will be phenomenologically viable. The following sections detail the conditions on the GGSO matrix elements required for states to remain in the model. Randomly sampling GGSO matrices allows us to find the frequency of models that satisfy the conditions.

\section{Analysis of Tachyonic Sectors}
\label{sec: tachyons}
Tachyonic states with negative level-matched mass squared, $\mathcal{M}_{L}^{2} = \mathcal{M}^{2}_{R} = \mathcal{M}^{2} < 0 $, indicate the model lies on a maximum of the potential. In order to ensure that we find models that are phenomenologically viable and stable, we use the condition that all level-matched tachyons are projected from the spectrum, though tachyonic models that are stable can be constructed \cite{breitenlohner1982positive}. In this regard the sample could be considered to be overly constricted, however these conditions follow the precedent set in the literature \cite{faraggi2020towards,faraggi2021classification,faraggi2021satisfiability} and are chosen such that comparison to $S$--models is convenient. 
Without employing supersymmetry to stabilise our model, each tachyonic state must be accounted for and projected individually. A detailed review of the tachyonic sectors and their mass levels is given in \cite{faraggi2020towards}, which analyses a similar $\tilde{S}$--model, but with a slightly different basis set. For completeness, the on shell tachyons are defined in Table \ref{tab:my_table1}. Similarly, examples of the survival conditions of the tachyonic sectors are given in Table \ref{tab:my_table2} and Table \ref{tab:my_table3}. These conditions are derived from eq. (\ref{eq:Hil}).
\begin{table}[htbp]
\centering
\begin{tabular}{|c|c|c|}
\hline
Mass level & Vectorial States & Spinorial States \\\hline
&&\\
$(-\frac{1}{2}, -\frac{1}{2})$ & $ \bar\lambda_{m} | \mathbf{0} \rangle $ & $|\mathbf{z}_{1,2} \rangle$ \\ 
 &&\\
 \hline
 &&\\
$(-\frac{3}{8}, -\frac{3}{8})$ &  $\bar\lambda_{m} | \mathbf{e}_{i} \rangle$ &  $| \mathbf{e}_{i}+\mathbf{z}_{1,2} \rangle$\\ 
&&\\\hline
&&\\
$(-\frac{1}{4}, -\frac{1}{4})$ & $\bar\lambda_{m} | \mathbf{e}_{i}+\mathbf{e}_{j} \rangle$ & $| \mathbf{e}_{i}+\mathbf{e}_{j}+\mathbf{z}_{1,2} \rangle$\\ 
&&\\\hline
&&\\
$(-\frac{1}{8}, -\frac{1}{8})$ & $\bar\lambda_{m} | \mathbf{e}_{i}+\mathbf{e}_{j}+\mathbf{e}_{k} \rangle$ & $| \mathbf{e}_{i}+\mathbf{e}_{j}+\mathbf{e}_{k}+\mathbf{z}_{1,2} \rangle$\\
&&\\
\hline
\end{tabular}
\caption{\label{tab:my_table1} Level-matched tachyonic states and their corresponding mass level, with $i \neq j \neq k = 1,...,6$ and $\bar\lambda_{m}$ being any right-moving complimentary fermion oscillator}
\end{table}
A key feature of this model is the relationship between the tachyonics states and the observable enhancements. In particular the survival conditions for the $|\bm{z}_{1,2} \rangle$ tachyonic states are the same as those for the $\psi^{\mu} \bar{\psi}^{1,...,5(*)}|\bm{z}_{1,2} \rangle$ enhancements to the gauge group. Therefore, for this choice of basis vectors, there are no tachyon free models with an enhanced $SO(10)$ symmetry. This is discussed further in Subsection \ref{subsec:Enhancements}.

\begin{table}[htbp]
\centering
\begin{tabular}{|c|c|}
\hline
Spinorial State & Survival Condition \\ \hline
$\mathbf{z}_{1}$ & $C\begin{bmatrix} \mathbf{z}_{1} \\ \mathbf{e}_{1,...,6} \end{bmatrix} =C\begin{bmatrix} \mathbf{z}_{1} \\ \mathbf{b}_{1,2} \end{bmatrix} =C\begin{bmatrix} \mathbf{z}_{1} \\ \mathbf{z}_{2} \end{bmatrix} = +1$ \\ \hline
$\mathbf{z}_{2}$ & $C\begin{bmatrix} \mathbf{z}_{2} \\ \mathbf{e}_{1,...,6} \end{bmatrix} =C\begin{bmatrix} \mathbf{z}_{2} \\ \mathbf{b}_{1,2} \end{bmatrix} =C\begin{bmatrix} \mathbf{z}_{2} \\ \mathbf{z}_{1} \end{bmatrix} = +1$ \\ \hline
$\mathbf{e}_{1}+ \mathbf{z}_{1}$ & $C\begin{bmatrix} \mathbf{e}_{1}+ \mathbf{z}_{1} \\ \mathbf{e}_{2,...,6} \end{bmatrix} =C\begin{bmatrix} \mathbf{e}_{1}+ \mathbf{z}_{1} \\ \mathbf{b}_{1} \end{bmatrix} =C\begin{bmatrix} \mathbf{e}_{1}+ \mathbf{z}_{1}\\ \mathbf{z}_{2} \end{bmatrix} = +1$ \\ \hline
$\mathbf{e}_{1}+ \mathbf{z}_{2}$ & $C\begin{bmatrix} \mathbf{e}_{1}+ \mathbf{z}_{2} \\ \mathbf{e}_{2,...,6} \end{bmatrix} =C\begin{bmatrix} \mathbf{e}_{1}+ \mathbf{z}_{2} \\ \mathbf{b}_{1} \end{bmatrix} =C\begin{bmatrix} \mathbf{e}_{1}+ \mathbf{z}_{2}\\ \mathbf{z}_{1} \end{bmatrix} = +1$ \\ \hline

$\mathbf{e}_{1}+\mathbf{e}_{2}+ \mathbf{z}_{1}$ & $C\begin{bmatrix} \mathbf{e}_{1}+\mathbf{e}_{2}+ \mathbf{z}_{1} \\ \mathbf{e}_{3,...,6} \end{bmatrix} =C\begin{bmatrix} \mathbf{e}_{1}+\mathbf{e}_{2}+ \mathbf{z}_{1} \\ \mathbf{b}_{1} \end{bmatrix} =C\begin{bmatrix} \mathbf{e}_{1}+\mathbf{e}_{2}+ \mathbf{z}_{1}\\ \mathbf{z}_{2} \end{bmatrix} = +1$ \\ \hline
$\mathbf{e}_{1}+\mathbf{e}_{2}+ \mathbf{z}_{2}$ & $C\begin{bmatrix} \mathbf{e}_{1}+\mathbf{e}_{2}+ \mathbf{z}_{2} \\ \mathbf{e}_{3,...,6} \end{bmatrix} =C\begin{bmatrix} \mathbf{e}_{1}+\mathbf{e}_{2}+ \mathbf{z}_{2} \\ \mathbf{b}_{1} \end{bmatrix} =C\begin{bmatrix} \mathbf{e}_{1}+\mathbf{e}_{2}+ \mathbf{z}_{2}\\ \mathbf{z}_{1} \end{bmatrix} = +1$ \\ \hline
$\mathbf{e}_{1}+\mathbf{e}_{2}+\mathbf{e}_{3}+ \mathbf{z}_{1}$ & $C\begin{bmatrix} \mathbf{e}_{1}+\mathbf{e}_{2}+\mathbf{e}_{3}+ \mathbf{z}_{1} \\ \mathbf{e}_{4,5,6} \end{bmatrix} =C\begin{bmatrix} \mathbf{e}_{1}+\mathbf{e}_{2}+\mathbf{e}_{3}+ \mathbf{z}_{1}\\ \mathbf{z}_{2} \end{bmatrix} = +1$ \\ \hline
$\mathbf{e}_{1}+\mathbf{e}_{2}+\mathbf{e}_{3}+ \mathbf{z}_{2}$ & $C\begin{bmatrix} \mathbf{e}_{1}+\mathbf{e}_{2}+\mathbf{e}_{3}+ \mathbf{z}_{2} \\ \mathbf{e}_{4,5,6} \end{bmatrix} =C\begin{bmatrix} \mathbf{e}_{1}+\mathbf{e}_{2}+\mathbf{e}_{3}+ \mathbf{z}_{2}\\ \mathbf{z}_{1} \end{bmatrix} = +1$ \\ \hline

\end{tabular}
\caption{\label{tab:my_table2} Level-matched tachyonic spinorial states and the survival conditions of their GGSO Coefficients. This is not an exhaustive list and analogous conditions were found for other combinations of $\mathbf{e}_{i}$.}
\end{table}
\begin{table}[htbp]
\centering
\resizebox{\textwidth}{!}{%
\begin{tabular}{|c|c|c|c|c|c|c|c|c|c|}
\hline
Vectorial State & $C\begin{bmatrix} \mathbf{e}_{1}\\ \bm{\tilde{S}} \end{bmatrix}$ & $C\begin{bmatrix} \mathbf{e}_{1}\\ \mathbf{e}_{2} \end{bmatrix}$ & $C\begin{bmatrix} \mathbf{e}_{1}\\ \mathbf{e}_{3} \end{bmatrix}$ & $C\begin{bmatrix} \mathbf{e}_{1}\\ \mathbf{e}_{4} \end{bmatrix}$ & $C\begin{bmatrix} \mathbf{e}_{1}\\ \mathbf{e}_{5} \end{bmatrix}$& $C\begin{bmatrix} \mathbf{e}_{1}\\ \mathbf{e}_{6} \end{bmatrix}$& $C\begin{bmatrix} \mathbf{e}_{1}\\ \mathbf{b}_{1} \end{bmatrix}$ & $C\begin{bmatrix} \mathbf{e}_{1}\\ \mathbf{z}_{1} \end{bmatrix}$& $C\begin{bmatrix} \mathbf{e}_{1}\\ \mathbf{z}_{2} \end{bmatrix}$\\ \hline
$\bar{y_{2}}|\mathbf{e}_{1} \rangle $ &+ & - & + & + & + & + & - & + & + \\
\hline
$\bar{y_{3}}|\mathbf{e}_{1} \rangle $ &+ & + & - & + & + & + & - & + & + \\
\hline
$\bar{y_{4}}|\mathbf{e}_{1} \rangle $ &+ & + & + & - & + & + & - & + & + \\
\hline
$\bar{y_{5}}|\mathbf{e}_{1} \rangle $ &+ & + & + & + & - & + & - & + & + \\
\hline
$\bar{y_{6}}|\mathbf{e}_{1} \rangle $ &+ & + & + & + & + & - & - & + & + \\
\hline
$\bar{w_{2}}|\mathbf{e}_{1} \rangle $ &+ & - & + & + & + & + & + & + & + \\
\hline
$\bar{w_{3}}|\mathbf{e}_{1} \rangle $ &+ & + & - & + & + & + & + & + & + \\
\hline
$\bar{w_{4}}|\mathbf{e}_{1} \rangle $ &+ & + & + & - & + & + & + & + & + \\
\hline
$\bar{w_{5}}|\mathbf{e}_{1} \rangle $ &+ & + & + & + & - & + & + & + & + \\
\hline
$\bar{w_{6}}|\mathbf{e}_{1} \rangle $ &+ & + & + & + & + & - & + & + & + \\
\hline
$ \bar{\psi^{1,...,5(*)}} / \bar{\eta^{1(*)}}|\mathbf{e}_{1} \rangle $ & + & + & + & + & + & + & - & + & + \\
\hline
$ \bar{\eta^{2,3(*)}}|\mathbf{e}_{1} \rangle $ & + & + & + & + & + & + & + & + & + \\
\hline
$ \bar{\phi^{1,2(*)}} |\mathbf{e}_{1} \rangle $ & + & + & + & + & + & + & + & - & + \\
\hline
$ \bar{\phi^{3,4(*)}} |\mathbf{e}_{1} \rangle $ & - & + & + & + & + & + & + & - & + \\
\hline
$ \bar{\phi^{5,6(*)}} |\mathbf{e}_{1} \rangle $ & - & + & + & + & + & + & + & + & - \\
\hline
$ \bar{\phi^{7,8(*)}} |\mathbf{e}_{1} \rangle $ & + & + & + & + & + & + & + & + & - \\
\hline
\end{tabular}
}
\caption{\label{tab:my_table3} Level-matched tachyonic vectorial states and the survival conditions of their GGSO Coefficients. This is not an exhaustive list but demonstrates the trend.}
\end{table}
As demonstrated in Tables \ref{tab:my_table2} and \ref{tab:my_table3}, the spinorial tachyons have slightly simpler conditions for survival than their vectorial counterpart, and this is due to the presence of the oscillator. All possible oscillators must be accounted for and projected. Once all $126$ possible tachyonic states have been projected, we can build up the massless spectrum. 

\section{Analysis of Massless Sectors}
\label{sec: Massless}
Having removed level-matched tachyonic states from the spectrum, we can turn our attention to the massless states. It is from these states that
we can build our $SO(10)$ GUT model, and so we must ensure that they are compatible with the observed SM data. The key states we will look for are the fermion producing spinorial $\mathbf{16/\overline{16}}$ representations, and the boson producing vectorial $\mathbf{10}$ representations.
\subsection{Spinorials}
\label{subsec: Spinorial}
The spinorial $\boldsymbol{ 16 /\overline{16} }$ representation of $SO(10)$ appear as the following 48 states: 
\begin{eqnarray}
    \bm{F}^{(1)}_{pqrs} & = & \mathbf{b}_{1} + p \mathbf{e}_{3} + q \mathbf{e}_{4} + r \mathbf{e}_{5} + s \mathbf{e}_{6} 
    \label{eq:B1}\\
& = &
\{\psi^{\mu}, \chi^{1,2},(1-p)y^{3}\bar y^{3},pw^{3}\bar w^{3},(1-q)y^{4}\bar y^{4},qw^{4}\bar w^{4},\nonumber\\
& & ~~~~~~~~~~~~~
(1-r)y^{5}\bar y^{5},rw^{5}\bar w^{5},(1-s)y^{6}\bar y^{6},sw^{6}\bar w^{6}, \bar \eta^{1}, \bar \psi^{1,...,5}\}\nonumber\\
 \bm{F}^{(2)}_{pqrs} & = & \mathbf{b}_{2} + p \mathbf{e}_{1} + q \mathbf{e}_{2} + r \mathbf{e}_{5} + s \mathbf{e}_{6} 
     \label{eq:B2}\\
\bm{F}^{(3)}_{pqrs} & = & \mathbf{b}_{3} + p \mathbf{e}_{1} + q \mathbf{e}_{2} + r \mathbf{e}_{3} + s \mathbf{e}_{4} 
   \label{eq:B3}
 \end{eqnarray}


where $pqrs = 0,1$.
Here the vector $\mathbf{b_{3}} $ is constructed from the following basis vectors:
\begin{equation}
\begin{split}
\label{eq:b3}
    \mathbf{b}_{3} &= \one + \mathbf{b}_{1} + \mathbf{b}_{2} +\mathbf{z}_{1}+\mathbf{z}_{2} \\
    &=\{\psi^{\mu}, \chi^{5,6}, w^{1,2}, w^{3,4} | \bar w^{1,2}, \bar w^{3,4}, \eta^{3} \bar \psi^{1,...,5}\}
\end{split}
    \end{equation}
In order to count the number of $\boldsymbol{16}/\boldsymbol{\overline{16}}$ in the model, we can define the projectors, $P^{(i)}_{pqrs}$, and the chiral phases, $X^{(i)}_{pqrs}$, of each of these states.
The projectors for these sectors are as follows:

\begin{equation}
    \label{eq:P1}
P^{(1)}_{pqrs} = \frac{1}{2^{4}} \prod_{i=1,2} \left( 1-C\begin{bmatrix}
\bm{F}^{(1)}_{pqrs} \\
\mathbf{e}_{i}  \\
\end{bmatrix}^{*} \right) 
\prod_{a=1,2} \left( 1-C\begin{bmatrix}
\bm{F}^{(1)}_{pqrs} \\
\mathbf{z}_{a}  \\

\end{bmatrix}^{*} \right)
\end{equation}
\begin{equation}
    \label{eq:P2}
P^{(2)}_{pqrs} = \frac{1}{2^{4}} \prod_{i=3,4} \left( 1-C\begin{bmatrix}
\bm{F}^{(2)}_{pqrs} \\
\mathbf{e}_{i}  \\
\end{bmatrix}^{*} \right) 
\prod_{a=1,2} \left( 1-C\begin{bmatrix}
\bm{F}^{(2)}_{pqrs} \\
\mathbf{z}_{a}  \\

\end{bmatrix}^{*} \right)
\end{equation}

\begin{equation}
    \label{eq:P3}
P^{(3)}_{pqrs} = \frac{1}{2^{4}} \prod_{i=5,6} \left( 1-C\begin{bmatrix}
\bm{F}^{(3)}_{pqrs} \\
\mathbf{e}_{i}  \\
\end{bmatrix}^{*} \right) 
\prod_{a=1,2} \left( 1-C\begin{bmatrix}
\bm{F}^{(3)}_{pqrs} \\
\mathbf{z}_{a}  \\

\end{bmatrix}^{*} \right)
\end{equation}
Similarly, the chiral phases are the following:
\begin{equation}
    \label{eq:X1}
X^{(1)}_{pqrs} = -C\begin{bmatrix}
\bm{F}^{(1)}_{pqrs} \\
\mathbf{b}_{2}+r \mathbf{e}_{5}+ s \mathbf{e}_{6}  \\
\end{bmatrix}^{*} 
\end{equation}
\begin{equation}
    \label{eq:X2}
X^{(2)}_{pqrs} = -C\begin{bmatrix}
\bm{F}^{(2)}_{pqrs} \\
\mathbf{b}_{1}+ r \mathbf{e}_{5}+ s \mathbf{e}_{6}  \\
\end{bmatrix}^{*} 
\end{equation}
\begin{equation}
    \label{eq:X3}
X^{(3)}_{pqrs} = -C\begin{bmatrix}
\bm{F}^{(3)}_{pqrs} \\
\mathbf{b}_{1}+r \mathbf{e}_{3}+s \mathbf{e}_{4}  \\
\end{bmatrix}^{*} 
\end{equation}

From these, we can determine the numbers $N_{16}$ and  $N_{\overline{16}}$ of 
$\boldsymbol{16}/\boldsymbol{\overline{16}}$ $SO(10)$ representations, 
\begin{equation}
    \label{eq:N16}
N_{16} = \frac{1}{2} \sum_{\substack{A=1,2,3 \\ pqrs=0,1}} P^{A}_{pqrs} (1+X^{A}_{pqrs})
\end{equation}
\begin{equation}
    \label{eq:N16bar}
    N_{\overline{16}} = \frac{1}{2} \sum_{\substack{A=1,2,3 \\ pqrs=0,1}} P^{A}_{pqrs} (1-X^{A}_{pqrs})
    \end{equation}

To ensure the model is compatible with the $3$ generational SM, we must enforce the constraint:
\begin{equation}\label{Fert16s}
N_{16} - N_{\overline{16}} \geq 6
\end{equation}
This ensures that the model has 6 or more chiral generations at the $SO(10)$ level and is a necessary but not sufficient condition on the presence of three generations at the subgroup level. For example, additional basis vectors are required to break to the group down to Pat--Salam 
\cite{pati1974lepton, antoniadis1990three, acfkr2} or 
Standard--like Models \cite{fny,slm1, SLMclass} and their projection conditions will determine whether $3$ chiral generations remain in the spectrum. 
We remark that Pati--Salam ${\tilde S}$--models are not viable due to the 
absence from the spectrum
of the heavy Higgs states required to break the Pati--Salam 
symmetry down to the Standard Model, whereas the Standard--like 
${\tilde S}$--models may be viable \cite{faraggi2021classification}.
%
In contrast to the supersymmetric $S$--Models, there are no bosonic superpartners to these spinorial $\boldsymbol{16}/\boldsymbol{\overline{16}}$'s. In particular, adding $\bm{\tilde{S}}$ to $\bm{F}^{(i)}_{pqrs}$ states produces massive states coupled to the hidden sector.
We note that in the free fermionic heterotic--string models with an 
unbroken GUT symmetry, these sectors give rise to the heavy Higgs states 
that are used to break the GUT symmetry.

\subsection{Vectorials}
\label{sec:Vectors}

To map from the spinorial to the vectorial sectors, we define the vector $\tilde{\mathbf{x}}$ 
\begin{equation}
\begin{split}\label{eq:xtilde}
& \tilde{\mathbf{x}} = \sum_{i=1,2,3} \mathbf{b}_{i} + \sum_{a=1,2,3,4,5,6} \mathbf{e}_{i} \\
&\quad = \{ \psi^{\mu}, \chi^{1,2,3,4,5,6} ~|~ \bar \eta^{1,2,3}, \bar \psi^{1,...,5}\}.
\end{split}
\end{equation}
This sector maps the spacetime fermions in the spinorial $\boldsymbol{16}/\boldsymbol{\overline{16}}$ representation of $SO(10)$ to sectors that produce spacetime bosons in its vectorial representation. In other words SM matter particle states map to the Standard Model Higgs-compatible states. 
 
The vectorial $\boldsymbol{10}$ representation of $SO(10)$ 
is produced from the oscillators $\bar{\psi}^{a=1,...,5 (*)} $ acting in the sectors given by

\begin{align}
    \mathbf{V}^{(1)}_{pqrs} &= \bm{F}^{(1)}_{pqrs} + \tilde{\mathbf{x}} \notag \\
    &= \mathbf{b}_{2} + \mathbf{b}_{3} + p \mathbf{e}_{3} + q \mathbf{e}_{4} + r \mathbf{e}_{5} + s \mathbf{e}_{6} \notag \\
    &= \{\chi^{3,4}, \chi^{5,6},(1-p)y^{3}\bar{y}^{3},pw^{3}\bar{w}^{3},(1-q)y^{4}\bar{y}^{4},qw^{4}\bar{w}^{4}, \notag \\
    & \quad \quad(1-r)y^{5}\bar{y}^{5},rw^{5}\bar{w}^{5},(1-s)y^{6}\bar{y}^{6},sw^{6}\bar{w}^{6}, \bar{\eta}^{2,3}\} \label{eq:V1} \\
    \mathbf{V}^{(2)}_{pqrs} &= \bm{F}^{(2)}_{pqrs} + \tilde{\mathbf{x}} \label{eq:V2} \\
    \mathbf{V}^{(3)}_{pqrs} &= \bm{F}^{(3)}_{pqrs} + \tilde{\mathbf{x}} \label{eq:V3}
\end{align}
     
    Again we can define projectors, $R^{(i)}_{pqrs}$, to determine if these states remain in the spectrum
    \begin{equation}
    \label{eq:R1}
R^{(1)}_{pqrs} = \frac{1}{2^{4}} \prod_{i=1,2} \left( 1+C\begin{bmatrix}
\mathbf{e}_{i}  \\
\mathbf{V}^{(1)}_{pqrs} \\
\end{bmatrix}^{*} \right) 
\prod_{a=1,2} \left( 1+C\begin{bmatrix}
\mathbf{z}_{a}  \\
\mathbf{V}^{(1)}_{pqrs} \\
\end{bmatrix}^{*} \right)
\end{equation}
\begin{equation}
    \label{eq:R2}
R^{(2)}_{pqrs} = \frac{1}{2^{4}} \prod_{i=3,4} \left( 1+C\begin{bmatrix}
\mathbf{e}_{i}  \\
\mathbf{V}^{(2)}_{pqrs} \\
\end{bmatrix}^{*} \right) 
\prod_{a=1,2} \left( 1+C\begin{bmatrix}
\mathbf{z}_{a}  \\
\mathbf{V}^{(2)}_{pqrs} \\
\end{bmatrix}^{*} \right)
\end{equation}
\begin{equation}
    \label{eq:R3}
R^{(3)}_{pqrs} = \frac{1}{2^{4}} \prod_{i=5,6} \left( 1+C\begin{bmatrix}
\mathbf{e}_{i}  \\
\mathbf{V}^{(3)}_{pqrs} \\
\end{bmatrix}^{*} \right) 
\prod_{a=1,2} \left( 1+C\begin{bmatrix}
\mathbf{z}_{a}  \\
\mathbf{V}^{(3)}_{pqrs} \\
\end{bmatrix}^{*} \right)
\end{equation}
and the number $N_{10}$ that remain can be written as
\begin{equation}
    \label{eq:N10}
N_{10} =  \sum_{\substack{A=1,2,3 \\ pqrs=0,1}} R^{A}_{pqrs}. 
\end{equation}
To make our model compatible with the SM Higgs, we must ensure that at least one vectorial state remains that can produce the Higgs doublet
\begin{equation}\label{VecFert}
    N_{10} \geq 1.
\end{equation}

\subsection{Hidden Sectors}
\label{subsec: Hidden}

Additional massless sectors will also contribute to the spectrum and partition function. A consequence of the $\bm{\tilde{S}}$ mapping is the production of large numbers of massless hidden sectors. These other massless groups take the general form:
\begin{equation}
\label{eq:gen_hid}
    \mathbf{H}^{a,b,c,d}_{pqrs} = V^{(a)}_{pqrs} + b \mathbf{z}_{1} +c \mathbf{z}_{2} +d \St 
\end{equation}
In particular the following 96 spinorial sectors will give rise to spacetime bosons:
\begin{equation}
\mathbf{H}^{(a), (i)}_{pqrs} = V^{(a)}_{pqrs} +\mathbf{z}_{i} 
\end{equation}
And the 192 following sectors giving hidden spacetime fermions:
\begin{equation}
\mathbf{H}^{a,b,c, 1}_{pqrs} = V^{(a)}_{pqrs}  +b \mathbf{z}_{1} + c \mathbf{z}_{2}+\St
\end{equation}
with $a=1,2,3$, $b,c,d = 0,1 $, $pqrs = 0,1$  and $i=1,2$. 
The presence or absence of these hidden sectors will effect the constant (massless) term in the partition function $q$ expansion and so has an effect on the cosmological constant of the model. Similarly, the vectors $\mathbf{\delta}^{1,...,30}$ contribute 30 hidden sectors which do not fit into the general structure given in eq. (\ref{eq:gen_hid}). 
\begin{equation}
\mathbf{\delta}^{1,...,30} = \mathbf{e}_{i}+\mathbf{e}_{j}+\mathbf{e}_{k}+\mathbf{e}_{l}+\mathbf{z}_{a}
\end{equation}
where $i \neq j \neq k \neq l  = 1, ... ,6$ and $a=1,2$.
\subsection{Other Massless States}
As the entire massless spectrum contributes to the partition function and potential, we briefly describe other contributions to the massless spectrum, which come in the following forms: \\
\begin{equation}
\mathbf{\gamma}^{1,...,15} = \mathbf{e}_{i}+\mathbf{e}_{j}+\mathbf{e}_{k}+\mathbf{e}_{l}  
\end{equation}
\begin{equation}
    \tilde{\mathbf{V}} =   \St + b \mathbf{z}_{1} + c \mathbf{z}_{2}
\end{equation}
where $i \neq j \neq k \neq l $, and $b,c = 0,1$

The $\tilde{\mathbf{x}}$ sector given in eq. (\ref{eq:xtilde}) will contribute to the observable sector in the form of additional $\boldsymbol{16/\overline{16}}$ spinorial representations of the $SO(10)$. Both $\gamma^{1,...,15}$ and $\tilde{\mathbf{V}}$ produce vectorial representations. Considering $\mathbf{\gamma}^{1,...,15}$, with oscillators $\bar{y}^{i}/\bar{w}^{i}/\bar{\eta}^{1,2,3}/\bar{\phi}^{1,...,8}$ these states are hidden sector states, however if the state has oscillator $\bar{\psi}^{1,...,5}$ these states will give vectorial $\boldsymbol{10}$ states.
Similarly, when $\tilde{\mathbf{V}}$ states have $y^{i}/w^{i}$ oscillators, the states contribute to the hidden sector, however if the oscillator is $\bar{\psi}^{1,...,5}/\bar{\eta}^{1,2,3}/\bar{\phi}^{1,...,8}$, with only complimentary NS $\bar{\phi}$ fermions, these states couple to both hidden and observable sectors. The vector--like states from these sectors can decouple from 
the chiral low scale spectrum by giving Vacuum Expectation Values to 
some fields in the massless string spectrum. 
These states do not couple in the same way as the previous $\mathbf{F}_{pqrs}^{i}$ and $\mathbf{V}_{pqrs}^{i}$, and so any compatible SM states will not be found in any of these additional massless states. 

\subsection{Top Quark Mass Coupling}
\label{subsec: TQMC}
So far none of the criteria implemented have anything to say about interaction terms of the theory. A minimum requirement for a SM compatible model is a mechanism to give mass to the heaviest fermion, the Top quark, by coupling it to the Higgs. 
In this respect we note that there are two types of couplings that can give rise to
top quark mass term, which involves twisted ${T}$, and untwisted ${U}$,  
matter fields. The first coupling is of the form $T_{i}T_{i}U_{i}$ where $i=1,2,3$ denotes the 
three twisted planes of the $\mathbb{Z}_2 \times \mathbb{Z}_2$ orbifolds. The second type of couplings is 
of the form $T_iT_jT_k$ with $i\ne j\ne k$. In the quasi--realistic models with unbroken Pati--Salam \cite{antoniadis1990three, acfkr2}, or 
Standard--like Model \cite{fny, slm1, SLMclass}, 
$SO(10)$ subgroup there exist a doublet--splitting mechanism \cite{dtsm}
that depends on the assignment of symmetric versus anti--symmetric boundary conditions for 
the set of internal worldsheet fermions that correspond to the compactified toroidal dimensions. 
Symmetric boundary conditions give rise to untwisted colour triplets, whereas anti--symmetric
boundary conditions give rise to electroweak doublets. 
Similarly, in vacua that contain a basis vector that
breaks the $SO(10)$ symmetry
to the flipped $SU(5)$ subgroup, there is a top--bottom quark 
Yukawa coupling selection rule \cite{yukawacoup}, 
where for symmetric boundary conditions 
a bottom--quark $T_iT_iU_i$ Yukawa coupling is selected 
and for asymmetric boundary conditions a top--quark 
Yukawa coupling is selected. 
In this paper the $SO(10)$ symmetry is unbroken and in principle both type of couplings are obtained. Given that the models that 
we consider use symmetric boundary conditions, we impose the presence of
the second type of couplings as a potential top quark mass term. 

We impose that the Top Quark Mass Coupling (TQMC) must also be preserved through the following conditions,  a detailed account of which is given in \cite{rizos2014top}. Preserving the TQMC requires us to fix certain GGSO phases, such that we retain two spinorial states and a Vectorial state (Higgs) that take part in this tree-level interaction. Without loss of generality we can choose the spinorials $\bm{F}^{(1)}_{0000} = \mathbf{b}_{1}$, $\bm{F}^{(2)}_{0000} = \mathbf{b}_{2}$ and vectorial  $\bm{F}^{(3)}_{1111} + \tilde{\mathbf{x}} = b_{3}+\mathbf{e}_{1}+\mathbf{e}_{2}+\mathbf{e}_{3}+\mathbf{e}_{4} + \tilde{\mathbf{x}}$. These states are chosen for ease of implementing the conditions. Having decided on the states, we find the following survival conditions: 

\begin{equation}
\label{eq:B^(1)}
C\begin{bmatrix}
\mathbf{e}_{i}  \\
\mathbf{b}_{1} \\
\end{bmatrix} = C\begin{bmatrix}
\mathbf{z}_{a} \\
\mathbf{b}_{1} \\
\end{bmatrix} = -1 \text{ for i =} 1,2 \text{ a =} 1,2
\end{equation}
\begin{equation}
\label{eq:B^(2)}
C\begin{bmatrix}
\mathbf{e}_{i}  \\
\mathbf{b}_{2} \\
\end{bmatrix} = C\begin{bmatrix}
\mathbf{z}_{a} \\
\mathbf{b}_{2} \\
\end{bmatrix} = -1 \text{ for i =} 3,4 \text{ a =} 1,2
\end{equation}

\begin{equation}
\label{eq:B^(3)}
C\begin{bmatrix}
\mathbf{z}_{i} \\
\mathbf{b}_{1} \\
\end{bmatrix}
C\begin{bmatrix}
\mathbf{z}_{i} \\
\mathbf{b}_{2} \\
\end{bmatrix}
C\begin{bmatrix}
\mathbf{z}_{i} \\
\mathbf{e}_{5} \\
\end{bmatrix}
C\begin{bmatrix}
\mathbf{z}_{i} \\
\mathbf{e}_{6} \\
\end{bmatrix}
= 1 \text{ for i =} 1,2
\end{equation}
\begin{equation}
\label{eq:B^(3)1}
C\begin{bmatrix}
\mathbf{e}_{i} \\
\mathbf{b}_{1} \\
\end{bmatrix}
C\begin{bmatrix}
\mathbf{e}_{i} \\
\mathbf{b}_{2} \\
\end{bmatrix}
C\begin{bmatrix}
\mathbf{e}_{i} \\
\mathbf{e}_{5} \\
\end{bmatrix}
C\begin{bmatrix}
\mathbf{e}_{i} \\
\mathbf{e}_{6} \\
\end{bmatrix}
= 1 \text{ for i =} 5,6
\end{equation}
\begin{equation}
C\begin{bmatrix}
\mathbf{b}_{1} \\
\mathbf{b}_{2} \\
\end{bmatrix} =C
\begin{bmatrix}
\mathbf{b}_{2} \\
\mathbf{b}_{1} \\
\end{bmatrix} = -1
\end{equation} 
Eq. (\ref{eq:B^(3)}) being somewhat redundant due to the previous tachyon projection conditions and eq. (\ref{eq:B^(1)}) and (\ref{eq:B^(2)}). 

\subsection{Enhancements}
\label{subsec:Enhancements}
Gauge enhancements arise within this class of models from the sectors
\begin{equation}
    \label{eq:enhance1}
\psi^{\mu} \{ \bar\lambda^{i} \} | \mathbf{z}_{1} \rangle
\end{equation}
\begin{equation}
    \label{eq:enhance2}
\psi^{\mu} \{ \bar\lambda^{i} \} | \mathbf{z}_{2} \rangle
\end{equation}
\begin{equation}
    \label{eq:enhance3}
\psi^{\mu} | \mathbf{z}_{1} + \mathbf{z}_{2} \rangle
\end{equation}
where $\bar\lambda^{i}$ are all possible right--moving NS oscillators. Survival conditions for the $|\mathbf{z}_{1} \rangle $ enhancements are listed in Table \ref{tab:Enhancements}. The observable enhancements to the $SO(10)$ gauge group arise for $\bgl^i = \bgps^{1,...,5}$ in the above sectors. 
Comparing Tables \ref{tab:Enhancements}  and \ref{tab:my_table2}, we see the survival conditions of these states are identical to the $| \mathbf{z}_{i} \rangle$ tachyonic states. Therefore these enhancements would necessarily be projected out when looking for stable models. 
\begin{table}[htbp]
\centering
\resizebox{\textwidth}{!}{%
\begin{tabular}{|c|c|c|c|c|c|c|c|c|c|}
\hline
Enhancement & $C\begin{bmatrix} \mathbf{z}_{1}\\ \mathbf{e}_{1} \end{bmatrix}$ & $C\begin{bmatrix} \mathbf{z}_{1}\\ \mathbf{e}_{2} \end{bmatrix}$ & $C\begin{bmatrix} \mathbf{z}_{1}\\ \mathbf{e}_{3} \end{bmatrix}$ & $C\begin{bmatrix} \mathbf{z}_{1}\\ \mathbf{e}_{4} \end{bmatrix}$ & $C\begin{bmatrix} \mathbf{z}_{1}\\ \mathbf{e}_{5} \end{bmatrix}$& $C\begin{bmatrix} \mathbf{z}_{1}\\ \mathbf{e}_{6} \end{bmatrix}$& $C\begin{bmatrix} \mathbf{z}_{1}\\ \mathbf{b}_{1} \end{bmatrix}$ & $C\begin{bmatrix} \mathbf{z}_{1}\\ \mathbf{b}_{2} \end{bmatrix}$& $C\begin{bmatrix} \mathbf{z}_{1}\\ \mathbf{z}_{2} \end{bmatrix}$\\ \hline
$\psi^{\mu} \bar{y_{1}}|\mathbf{z}_{1} \rangle $ &- & + & + & + & + & + & - & + & + \\
\hline
$\psi^{\mu} \bar{y_{2}}|\mathbf{z}_{1} \rangle $ &+ & - & + & + & + & + & - & + & + \\
\hline
$\psi^{\mu} \bar{y_{3}}|\mathbf{z}_{1} \rangle $ &+ & + & - & + & + & + & + & - & + \\
\hline
$\psi^{\mu} \bar{y_{4}}|\mathbf{z}_{1} \rangle $ &+ & + & + & - & + & + & + & - & + \\
\hline
$\psi^{\mu} \bar{y_{5}}|\mathbf{z}_{1} \rangle $ &+ & + & + & + & - & + & + & - & + \\
\hline
$\psi^{\mu} \bar{y_{6}}|\mathbf{z}_{1} \rangle $ &+ & + & + & + & + & - & + & - & + \\
\hline
$\psi^{\mu} \bar{w_{1}}|\mathbf{z}_{1} \rangle $ & - & + & + & + & + & + & - & - & + \\
\hline
$\psi^{\mu} \bar{w_{2}}|\mathbf{z}_{1} \rangle $ &+ & - & + & + & + & + & - & - & + \\
\hline
$\psi^{\mu} \bar{w_{3}}|\mathbf{z}_{1} \rangle $ &+ & + & - & + & + & + & - & - & + \\
\hline
$\psi^{\mu} \bar{w_{4}}|\mathbf{z}_{1} \rangle $ &+ & + & + & - & + & + & - & - & + \\
\hline
$\psi^{\mu} \bar{w_{5}}|\mathbf{z}_{1} \rangle $ &+ & + & + & + & - & + & - & + & + \\
\hline
$\psi^{\mu} \bar{w_{6}}|\mathbf{z}_{1} \rangle $ &+ & + & + & + & + & - & - & + & + \\
\hline
$\psi^{\mu}  \bar{\psi^{1,...,5(*)}} |\mathbf{z}_{1} \rangle $ & + & + & + & + & + & + & + & + & + \\
\hline
$\psi^{\mu}  \bar{\eta^{1(*)}}|\mathbf{z}_{1} \rangle $ & + & + & + & + & + & + & + & - & + \\
\hline
$\psi^{\mu}  \bar{\eta^{2(*)}}|\mathbf{z}_{1} \rangle $ & + & + & + & + & + & + & - & + & + \\
\hline
$\psi^{\mu}  \bar{\eta^{3(*)}}|\mathbf{z}_{1} \rangle $ & + & + & + & + & + & + & - & - & + \\
\hline
$\psi^{\mu}  \bar{\phi^{5,6(*)}} |\mathbf{z}_{1} \rangle $ & + & + & + & + & + & + & - & - & - \\
\hline
$\psi^{\mu}  \bar{\phi^{7,8(*)}} |\mathbf{z}_{1} \rangle $ & + & + & + & + & + & + & - & - & - \\
\hline
\end{tabular}
}
\caption{\label{tab:Enhancements} Enhancements and the survival conditions of their GGSO Coefficients, for the $| \mathbf{z}_{1} \rangle$ enhancements. The conditions for observable enhancemnets, $\psi^{\mu} \bar{\psi^{1,...,5(*)}}|\mathbf{z}_{1} \rangle$, are shown here to be the same as the condition for the tachyonic states.}
\end{table}

\subsection{$N_{b} - N_{f} = 0$ at the Massless Level}
It is possible to find models in which $N_{b} = N_{f}$ at the massless level as one would have for supersymmetric models at the free fermionic point (FFP). These models will appear sporadically as there is no inbuilt mechanism to produce this phenomenon. This condition could also be achieved by moving away from the FFP in moduli space \cite{florakis2017gravitational, florakis2016chiral,avalos2023d}. The typical motivation for this condition is in the context of `super no-scale' models \cite{Kounnas:2016gmz} where it plays a role in the exponential suppression of the one loop cosmological constant for models with spontaneous breaking of supersymmetry. Although the $\tilde{S}$--models exhibit explicit breaking of supersymmetry this Bose-Fermi degeneracy at the massless level is still an interesting feature of the spectrum to note due to its analogy to supersymmetric models.

\newpage 
\section{Partition Function and Potential}
\subsection{Partition Function}
\label{sec:Partfunc}
The partition function of a string model expresses information about the full tower of string states, both on--shell and off-shell. In particular, when written as a $q$--expansion, the total number of states at each mass can be read off. For example, the coefficient $a_{00} = N_{b} - N_{f}$ gives the net number of bosons and fermions at the massless level and so is necessarily $0$ in supersymmetric models. For non--supersymmetric models we typically get a much richer spectrum with $a_{ij} \neq 0$. One feature to emphasize is the off-shell tachyon contributions, including the model-independent`proto--graviton' state as demonstrated in ref. \cite{Dienes:1990ij}.

In the FFF, the partition function can be expressed as 
\begin{equation}
    \label{eq:Partfunctferm}
Z = Z_{B} \sum_{\Bga,\Bgb} \CC{\Bga}{\Bgb} \prod_f Z\begin{bmatrix} \mathbf{\alpha}(f) \\ \mathbf{\beta}(f) \end{bmatrix} 
\end{equation}
 where $\Bga$ and $\Bgb$ are sectors, the product is over the fermion boundary conditions in each sector and $Z_{B}$ is the spacetime boson contribution. However, it is more convenient to express the partition function in the following modular invariant way, as outlined in \cite{fknr, Viktorthesis, avalos2023d, FRtranslation,florakis2023free}:
 
\begin{equation}
\begin{split}
\label{eq:Z2} 
& Z = {\frac{1}{{\eta^{12}\bar\eta^{24}}}} \frac{1}{2^{3}} \sum_{\substack{a,k,\rho \\ b,l,\sigma}} 
\frac{1}{2^{6}} \sum_{\substack{\zeta_{i} \\ \delta_{i}}}
\frac{1}{2^{3}} \sum_{\substack{h_1,h_2,H \\ g_1,g_2,G}}
(-1)^{a+b+\phi \scalebox{0.6}{$ \begin{bmatrix} a&k&\rho&\zeta_{i}&h_{1}&h_{2}&H \\ b&l&\sigma&\delta_{i}&g_{1}&g_{2}&G \end{bmatrix}$}} \\ 
& \vartheta\begin{bmatrix} a \\b \end{bmatrix}_{\psi^{\mu}} 
\vartheta\begin{bmatrix} a+h_2 \\b+g_2 \end{bmatrix}_{\chi^{1,2}}
\vartheta\begin{bmatrix} a+h_1 \\b+g_1 \end{bmatrix}_{\chi^{3,4}} 
\vartheta\begin{bmatrix} a-h_1-h_2 \\b-g_1-g_2 \end{bmatrix}_{\chi^{5,6}} \\
&\vartheta\begin{bmatrix} \zeta_{1} \\ \delta_{1} \end{bmatrix}^{\frac{1}{2}}_{w^{1}}
\vartheta\begin{bmatrix} \zeta_{1}+h_2 \\ \delta_{1}+g_2 \end{bmatrix}^{\frac{1}{2}}_{y^{1}}
\bar\vartheta\begin{bmatrix} \zeta_{1} \\ \delta_{1} \end{bmatrix}^{\frac{1}{2}}_{\bar w^{1}}
\bar\vartheta\begin{bmatrix} \zeta_{1}+h_2 \\ \delta_{1}+g_2 \end{bmatrix}^{\frac{1}{2}}_{\bar y^{1}} \\
& \vartheta\begin{bmatrix} \zeta_{2} \\ \delta_{2} \end{bmatrix}^{\frac{1}{2}}_{w^{2}}
\vartheta\begin{bmatrix} \zeta_{2}+h_2 \\ \delta_{2}+g_2 \end{bmatrix}^{\frac{1}{2}}_{y^{2}}
\bar\vartheta\begin{bmatrix} \zeta_{2} \\ \delta_{2} \end{bmatrix}^{\frac{1}{2}}_{\bar w^{2}}
\bar\vartheta\begin{bmatrix} \zeta_{2}+h_2 \\ \delta_{2}+g_2 \end{bmatrix}^{\frac{1}{2}}_{\bar y^{2}} \\
&\vartheta\begin{bmatrix} \zeta_{3} \\ \delta_{3} \end{bmatrix}^{\frac{1}{2}}_{w^{3}}
\vartheta\begin{bmatrix} \zeta_{3}+h_1 \\ \delta_{3}+g_1 \end{bmatrix}^{\frac{1}{2}}_{y^{3}}
\bar\vartheta\begin{bmatrix} \zeta_{3} \\ \delta_{3} \end{bmatrix}^{\frac{1}{2}}_{\bar w^{3}}
\bar\vartheta\begin{bmatrix} \zeta_{3}+h_1 \\ \delta_{3}+g_1 \end{bmatrix}^{\frac{1}{2}}_{\bar y^{3}} \\
&\vartheta\begin{bmatrix} \zeta_{4} \\ \delta_{4} \end{bmatrix}^{\frac{1}{2}}_{w^{4}}
\vartheta\begin{bmatrix} \zeta_{4}+h_1 \\ \delta_{4}+g_1 \end{bmatrix}^{\frac{1}{2}}_{y^{4}}
\bar\vartheta\begin{bmatrix} \zeta_{4} \\ \delta_{4} \end{bmatrix}^{\frac{1}{2}}_{\bar w^{4}}
\bar\vartheta\begin{bmatrix} \zeta_{4}+h_1 \\ \delta_{4}+g_1 \end{bmatrix}^{\frac{1}{2}}_{\bar y^{4}} \\
& \vartheta\begin{bmatrix} \zeta_{5}+h_2 \\ \delta_{5}+g_2 \end{bmatrix}^{\frac{1}{2}}_{w^{5}}
\vartheta\begin{bmatrix} \zeta_{5}+h_1 \\ \delta_{5}+g_1 \end{bmatrix}^{\frac{1}{2}}_{y^{5}}
\bar\vartheta\begin{bmatrix} \zeta_{5}+h_2 \\ \delta_{5}+g_2 \end{bmatrix}^{\frac{1}{2}}_{\bar w^{5}}
\bar\vartheta\begin{bmatrix} \zeta_{5}+h_1 \\ \delta_{5}+g_1 \end{bmatrix}^{\frac{1}{2}}_{\bar y^{5}} \\
&\vartheta\begin{bmatrix} \zeta_{6}+h_2 \\ \delta_{6}+g_2 \end{bmatrix}^{\frac{1}{2}}_{w^{6}}
\vartheta\begin{bmatrix} \zeta_{6}+h_1 \\ \delta_{6}+g_1 \end{bmatrix}^{\frac{1}{2}}_{y^{6}}
\bar\vartheta\begin{bmatrix} \zeta_{6}+h_2 \\ \delta_{6}+g_2 \end{bmatrix}^{\frac{1}{2}}_{\bar w^{6}}
\bar\vartheta\begin{bmatrix} \zeta_{6}+h_1 \\ \delta_{6}+g_1 \end{bmatrix}^{\frac{1}{2}}_{\bar y^{6}} \\
&\bar{\vartheta}\begin{bmatrix} k \\ l \end{bmatrix}^{5}_{\bar\psi} 
\bar{\vartheta}\begin{bmatrix} k+h_2 \\ l+g_2 \end{bmatrix}_{\bar \eta^{1}} 
\bar{\vartheta}\begin{bmatrix} k+h_1 \\ l+g_1 \end{bmatrix}_{\bar \eta^{2}}
\bar{\vartheta}\begin{bmatrix} k-h_1-h_2 \\ l-g_1-g_2 \end{bmatrix}_{\bar \eta^{3}} \\
&\bar{\vartheta}\begin{bmatrix} \rho \\ \sigma \end{bmatrix}^{2} _{\bar\phi^{1,2}}
\bar{\vartheta}\begin{bmatrix} \rho + a + k \\ \sigma+b+l  \end{bmatrix}^{2} _{\bar\phi^{3,4}} 
\bar{\vartheta}\begin{bmatrix} \rho + a + k + H\\ \sigma+b+l+G  \end{bmatrix}^{2} _{\bar\phi^{5,6}}
\bar{\vartheta}\begin{bmatrix} \rho+H \\ \sigma+G \end{bmatrix}^{2} _{\bar\phi^{7,8}}
\end{split}
\end{equation}

In eq. (\ref{eq:Z2}), the modular invariant phase $\phi$ encodes the GGSO phase matrix and the spin statistics are encompassed in the phase $a+b$.
The indices $a,b$ are associated with the spacetime fermions $\psi^{\mu}$, $\zeta_{i}$ relate to the internal $12$ internal left and right moving degrees of freedom, $h_{1}$ and $h_{2}$ correspond to the orbifold twists and $k,l,\rho,\sigma, H, G $ relate to the $16$ additional right moving fermions.

We remark that while in all previous papers
the systematic analysis of the partition function and vacuum 
structure was applied exclusively to $S$--models, the analysis
in our paper 
is the first in which is it is applied to $\tilde{S}$--based models. That it should work is \textit{a priori} not at all guaranteed. We highlight that the key difference in the $\tilde{S}$--model is the appearance of indices $a, b$ in the last line of eq. (\ref{eq:Z2}), indicating the coupling of $\psi^{\mu}$ to the hidden sector fermions within $\bm{\tilde{S}}$.  
 The above eq. (\ref{eq:Z2}) is still only valid at the FFP and so if we allow ourselves to move in the moduli space, we need to rewrite the partition function accordingly.

For the $\mathbb{Z}_{2} \times \mathbb{Z}_{2}$ orbifold, the three internal tori associated with the $\Gamma_{2,2}$ lattice, are parametretised by three complex structure moduli and three K\"ahler moduli.
We can express the partition function in a way that makes the dependence on these geometric moduli, $T$ and $U$, manifest
\begin{equation}    
\begin{split}\label{eq:Z3}
& Z = \frac{1}{\eta^{12} \bar\eta^{24}} \frac{1}{2^{3}} \sum_{\substack{a,k,\rho \\ b,l,\sigma}}
\frac{1}{2^{6}} \sum_{\substack{\zeta_{i} \\ \delta_{i}}}
\frac{1}{2^{3}} \sum_{\substack{h_1,h_2,H \\ g_1,g_2,G}}
(-1)^{a+b+\phi \scalebox{0.6}{$\begin{bmatrix} {a}&{k}&{\rho}&{\zeta_{i}}&{h_{1}}&{h_{2}}&{H} \\ {b}&{l}&{\sigma}&{\delta_{i}}&{g_{1}}&{g_{2}}&{G} \end{bmatrix}$}} \\ 
& \vartheta\begin{bmatrix} a \\b \end{bmatrix}_{\psi^{\mu}}
\vartheta\begin{bmatrix} a+h_2 \\b+g_2 \end{bmatrix}_{\chi^{1,2}}
\vartheta\begin{bmatrix} a+h_1 \\b+g_1 \end{bmatrix}_{\chi^{3,4}} 
\vartheta\begin{bmatrix} a-h_1-h_2 \\b-g_1-g_2 \end{bmatrix}_{\chi^{5,6}} \\
& \Gamma^{1}_{2,2}
\left[
\begin{array}{c|c}
H_1 \text{ } H_2 & h_2 \\
G_1 \text{ } G_2 & g_2
\end{array}
\right](T^{(1)}*, U^{(1)}*)  \\
&\Gamma^{2}_{2,2}
\left[
\begin{array}{c|c}
H_3 \text{ } H_4 & h_1 \\
G_3 \text{ } G_4 & g_1
\end{array}
\right](T^{(2)}*, U^{(2)}*)  \\
&\Gamma^{3}_{2,2}
\left[
\begin{array}{c|c}
H_5 \text{ } H_6 & h_1 +  h_2 \\
G_5 \text{ } G_6 & g_1 +   g_2
\end{array}
\right](T^{(3)}*, U^{(3)}*)  \\
&\bar{\vartheta}\begin{bmatrix} k \\ l \end{bmatrix}^{5}_{\bar\psi} 
\bar{\vartheta}\begin{bmatrix} k+h_2 \\ l+g_2 \end{bmatrix}_{\bar \eta^{1}} 
\bar{\vartheta}\begin{bmatrix} k+h_1 \\ l+g_1 \end{bmatrix}_{\bar \eta^{2}}
\bar{\vartheta}\begin{bmatrix} k-h_1-h_2 \\ l-g_1-g_2 \end{bmatrix}_{\bar \eta^{3}} \\
&\bar{\vartheta}\begin{bmatrix} \rho \\ \sigma \end{bmatrix}^{2} _{\bar\phi^{1,2}}
\bar{\vartheta}\begin{bmatrix} \rho + a + k \\ \sigma+b+l  \end{bmatrix}^{2} _{\bar\phi^{3,4}} 
\bar{\vartheta}\begin{bmatrix} \rho + a + k + H\\ \sigma+b+l+G  \end{bmatrix}^{2} _{\bar\phi^{5,6}}
\bar{\vartheta}\begin{bmatrix} \rho+H \\ \sigma+G \end{bmatrix}^{2} _{\bar\phi^{7,8}} \text{ . }\\
\end{split}
\end{equation}
where, at the FFP,  $(T, U) = (T_{1} +i T_{2} , U_{1} +i U_{2}) = (T*, U*) = (i, \frac{i}{2}) $, and we define $\Gamma$ as:

\begin{equation}
\begin{split}\label{eq:gamma1}
&\Gamma_{2,2}
\left[
\begin{array}{c|c}
H_1 \text{ } H_2 & h_{2} \\
G_1 \text{ } G_2 & g_{2} \\
\end{array} \right] (i, \frac{i}{2})= \frac{1}{4} \sum_{\substack{\zeta_{i} \delta_{i} \in \mathbb{Z} }}  (-1)^{(\zeta_{i}+h_{2})G_{i} +(\delta_{i}+g_{2})H_{i} + H_{i}G_{i}} \\
&\vartheta\begin{bmatrix} \zeta_{1} \\ \delta_{1} \end{bmatrix}^{\frac{1}{2}}_{w^{1}}
\vartheta\begin{bmatrix} \zeta_{1}+h_2 \\ \delta_{1}+g_2 \end{bmatrix}^{\frac{1}{2}}_{y^{1}}
\bar\vartheta\begin{bmatrix} \zeta_{1} \\ \delta_{1} \end{bmatrix}^{\frac{1}{2}}_{\bar w^{1}}
\bar\vartheta\begin{bmatrix} \zeta_{1}+h_2 \\ \delta_{1}+g_2 \end{bmatrix}^{\frac{1}{2}}_{\bar y^{1}} \\
&\vartheta\begin{bmatrix} \zeta_{2} \\ \delta_{2} \end{bmatrix}^{\frac{1}{2}}_{w^{2}}
\vartheta\begin{bmatrix} \zeta_{2}+h_2 \\ \delta_{2}+g_2 \end{bmatrix}^{\frac{1}{2}}_{y^{2}}
\bar\vartheta\begin{bmatrix} \zeta_{2} \\ \delta_{2} \end{bmatrix}^{\frac{1}{2}}_{\bar w^{2}}
\bar\vartheta\begin{bmatrix} \zeta_{2}+h_2 \\ \delta_{2}+g_2 \end{bmatrix}^{\frac{1}{2}}_{\bar y^{2}} \text{ . }
\end{split}
\end{equation}
The dependence on the geometric moduli is contained in the untwisted sectors, i.e. the sectors where $h_{i} = g_{i} = 0$, so we also define the following:

\begin{equation}
\begin{split}
\label{eq:gamma2}
&\Gamma_{2,2}
\left[
\begin{array}{c|c}
H_1 \text{ } H_2 & 0 \\
G_1 \text{ } G_2 & 0
\end{array}
\right](T, U) = \sum_{\substack{m_{i} n_{i} \in \mathbb{Z} }} q^{\frac{1}{2} {|{P_{L}(T,U)|}^{2}}} \bar q^{\frac{1}{2} |{{P_{R}(T,U)|}^{2}}}e^{i \pi(G_{1}m_{1}+G_{2}n_{2})}\\
\end{split}
\end{equation} 
where
\begin{equation}
    \label{eq:PL}
P_{L} = \frac{1}{\sqrt{2T_{2}U_{2}}} \left[ m_2 +\frac{H_{2}}{2} - Um_1 +T(n_1 +\frac{H_1}{2} + Un_{2}) \right]
\end{equation}
\begin{equation}
    \label{eq:PR}
P_{R}  = \frac{1}{\sqrt{2T_{2}U_{2}}} \left[ m_2 +\frac{H_{2}}{2} - Um_1 +\bar{T}(n_1 +\frac{H_1}{2} + Un_{2}) \right]
\end{equation}
which is equal to the previous expression at the FFP $(T, U) = (T*, U*)$. Eqs. (\ref{eq:gamma2}), (\ref{eq:PL}) and (\ref{eq:PR} are independent of $a,b$ and so are identical in $S$-- and $\tilde{S}$--models. From this the $q$--expansion can be found and analysed, using standard definitions of $\vartheta$ and $\eta$ functions that can be found in Appendix A of \cite{faraggi2020towards}, for example. The general form of the $q$-expansion can be expressed as

\begin{equation}
    Z = \sum_{m,n} a_{m,n}q^{m}\bar{q}^n \text{.}
\end{equation}

\subsection{Potential}
\label{sec:Pot}
The resulting one-loop potential determines the stability of the model, with the ideal being a consistent theory sitting in a global minima of the potential.
Having determined a method of finding the partition function at general locations in the moduli space, we can observe how the one loop potential varies over the moduli space by integrating the partition function over the fundamental domain of the torus amplitude:
\begin{equation}
    \label{eq:Pot}
    \begin{split}
V_{1-loop} &= - \frac{1}{2} \frac{\mathcal{M}^{4}}{(2\pi)^{4}} \int_{\mathcal{F}} \frac{d^{2}\tau}{\tau^{2}_{3}} Z(\tau, \bar\tau ; T^{(i)}, U^{(i)})\\
&= - \frac{1}{2} \frac{\mathcal{M}^{4}}{(2\pi)^{4}} \int_{\mathcal{F}} \frac{d^{2}\tau}{\tau^{2}_{3}} \sum a_{m,n}q^{m}\bar{q}^n\\
&= - \frac{1}{2} \frac{\mathcal{M}^{4}}{(2\pi)^{4}} \sum a_{m,n} I_{m,n}
    \end{split}
\end{equation}
where $\frac{d^{2}\tau}{\tau^{2}_{2}}$
is the modular invariant measure and the fundamental domain, $\mathcal{F}$, is defined as
\begin{equation}
    \mathcal{F} = \{ \tau \in \mathbb{C} \text{ }| \quad |\tau|^{2} > 1 \quad \land \quad |\tau_{1}| < 0.5 \}\text{.}
\end{equation}
This corresponds to the spacetime cosmological constant of the theory, with the factor $- \frac{1}{2} \frac{\mathcal{M}^{4}}{(2\pi)^{4}}$ relating the worldsheet potential to the spacetime potential.  For example, supersymmetric models with $a_{ij} = 0$ have a vanishing cosmological constant. 

Whilst all of the models that fulfil the phenomenological criteria are found to have finite potential, only a fraction are stable at the FFP, as there is no inherent reason for the minima to be located here. This feature is a result of $\mathbf{e_{i}}$ basis vectors which generally break $T$-duality. In contrast, models that use $\mathbf{T_{i}}$ basis vectors \cite{florakis2016chiral} appear to always sit in minima at the FFP.
Moving away from the FFP typically leads to divergences. These divergences arise from level-matched and non-level matched tachyons, which can be understood through the modular integral results
\begin{equation}
\label{eq:inf}
I_{m,n} = 
\begin{cases} 
    \infty & \text{if } m+n < 0 \text{ and } m-n \not\in \mathbb{Z} \setminus \{0\} \\
    \text{Finite} & \text{otherwise}.
\end{cases}
\end{equation}
With $T$ and $U$ being complex moduli, we have $6$ complex internal degrees of freedom. It is impractical to vary all $6$ and determine the shape of the potential. Following the logic used in refs. 
\cite{florakis2016chiral, avalos2023d, florakis2022super}, 
we choose to vary: $\Im(T^{(1)}) = T_{2}$ associated with the volume of the first torus and $\Im(U^{(1)}) = U_{2}$, the imaginary component of $U^{(1)}$, associated with the shape, while leaving the moduli of the second and third tori fixed at the free fermionic point. The modulus $T_{2}$ has previously been studied in relation to explicit and Scherk-Schwarz spontaneous supersymmetry breaking \cite{avalos2023d}, where, in the latter case, SUSY can be restored as $T_{2} \rightarrow \infty$. We have chosen here to vary the same moduli for consistency and comparison between $S$--models and our $\tilde{S}$--models.

\section{Classification Results}
\label{sec:Results}
Now that we have defined our classification constraints we can classify models within the full space of $2^{66}\sim 7.4 \times 10^{19}$ GGSO phase configurations. Due to constraints on computational power and time, we take a sample size of $10^{9}$ models, which demonstrates the statistics well. The results of this search are given in Table \ref{tab:freq}. Of the sampled models, only $0.531 \% $ were free from level-matched tachyons and we recall that this constraint necessitates that observable enhancements are absent. We then choose to project hidden enhancements leaving $0.487 \%$ of the sample.
Of the remaining models, $0.00308 \%$ of the sample satisfy the constraint on the net chirality of spinorial $\boldsymbol{16/\overline{16}}$'s of eq. (\ref{Fert16s}). This further reduces to $0.00307 \%$  when accommodating for the required vectorial $\boldsymbol{10}$ of eq. (\ref{VecFert}).
Finally, we see that $84$ models in our sample obey the TQMC conditions and so we do produce viable, fertile models at a frequency of $8.40 \times 10^{-8}$. Although the condition $a_{00} =N_b-Nf= 0$ is not a phenomenological requirement, we show that in our sample $4$ models were found that obey this at the FFP.

\begin{table}[h]
\centering
\begin{tabular}{|c|c|c|}
\hline
Constraint & Quantity & Probability \\\hline
No Constraints & $1.00 \times 10^{9}$ & 1 \\
Tachyon Free &  $ 5,309,285$ & $5.309 \times 10^{-3}$\\
No Observable Enhancements & $ 5,309,285 $ & $5.309 \times 10^{-3}$\\
No Hidden Enhancements & $ 4,865,203 $ & $4.865 \times 10^{-3}$\\
$N_{16} - N_{\overline{16}} \geq 6$ & $ 30,773$  & $ 3.077 \times 10^{-5}$ \\
$N_{10} \geq 1$ & $ 30,717$ & $ 3.072 \times 10^{-5}$ \\
TQMC & $84$ & $8.40 \times 10^{-8}$ \\
$a_{00} = N^{(0)}_{b} - N^{(0)}_{f} = 0$ & $4$ & $4.00 \times 10^{-9}$\\
\hline
\end{tabular}
\caption{\label{tab:freq} Frequency of models after each additional constraint is applied to the sample of $1.00 \times 10^{9}$}
\end{table}

Further analysis of these models shows that they all have finite potential at the FFP, confirming the absence of level matched tachyons. As there are no complex boundary conditions in our models, non-level matched tachyons that would cause divergences are absent from the spectrum at the FFP. However the FFP is not necessarily the minimum in potential. This is demonstrated in the next section. 

\begin{figure}
    \centering
    \includegraphics[width=0.95\textwidth]{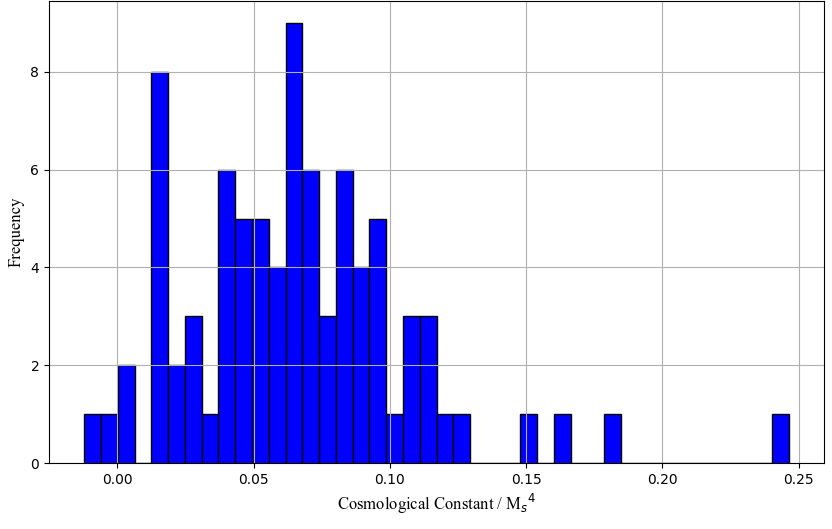}
    \caption{Histograph to show the distribution of the Cosmological Constant of the 84 fertile cores at the Free Fermionic Point.}
    \label{fig:hist}
\end{figure}

We mostly find models with positive values of the potential at the one loop level in this sample, representing De Sitter vacua. The distribution of the Cosmological Constants at the FFP is given in \ref{fig:hist}.  Only $2$ of the $84$ fertile cores have negative cosmological constant at the FFP, and these $2$ very quickly become divergent away from this point. $\tilde{S}$--models are more inclined to positive values of the potential. A significant factor in this is the projection of massive bosonic superpartners to higher masses, reducing their contribution to the potential.

\section{Example Model}
\label{sec:Ex}
Here we present an example of a model that meets all of the criteria in Table \ref{tab:freq}, save for $a_{00} = 0$. Below is the GGSO matrix of the model, from which we can derive the partition function.

\begin{equation}
C \begin{blockarray}{c}
\begin{block}{[c]}
    v_{i} \\
    v_{j} \\
\end{block}
\end{blockarray} =  \text{  }  
\begin{blockarray}{ccccccccccccc}
& \mathds{1} & \bm{\tilde{S}} & \mathbf{e}_{1} & \mathbf{e}_{2} & \mathbf{e}_{3} & \mathbf{e}_{4} & \mathbf{e}_{5} & \mathbf{e}_{6} & \mathbf{b}_{1} & \mathbf{b}_{2} & \mathbf{z}_{1} & \mathbf{z}_{2}\\
\begin{block}{c(cccccccccccc)}
\mathds{1} & -1 &~~1 &~~1 &-1 &~~1 &~~1 &-1 &~~1 &~~1 &~~1 &~~1 &~~1 \\
\bm{\tilde{S}} & ~~1 &-1 &-1 &~~1 &~~1 &~~1 &-1 &-1 &~~1 &~~1 &-1 &~~1 \\
\mathbf{e}_{1} & ~~1 &-1 &-1 &-1 &~~1 &~~1 &-1 &~~1 &-1 &~~1 &~~1 &~~1\\
\mathbf{e}_{2} & -1 &~~1 &-1 &~~1 &-1 &-1 &~~1 &-1 &-1 &~~1 &-1 &-1\\
\mathbf{e}_{3} & ~~1 &~~1 &~~1 &-1 &-1 &~~1 &~~1 &~~1 &~~1 &-1 &-1 &~~1 \\
\mathbf{e}_{4} & ~~1 &~~1 &~~1 &-1 &~~1 &-1 &~~1 &~~1 &~~1 &-1 &-1& ~~1\\
\mathbf{e}_{5} & -1 &-1 &-1 &~~1 &~~1 &~~1 &~~1 &~~1 &-1 &-1& ~~1 &-1\\
\mathbf{e}_{6} &~~1 &-1 &~~1 &-1 &~~1 &~~1 &~~1 &-1& ~~1 &-1 &~~1 &-1\\
\mathbf{b}_{1} & ~~1 &-1 &-1 &-1 &~~1 &~~1 &-1 &~~1 &~~1 &-1 &-1 &-1\\
\mathbf{b}_{2} & ~~1 &-1 &~~1 &~~1 &-1 &-1 &-1 &-1 &-1 &~~1 &-1 &-1\\
\mathbf{z}_{1} &~~1 &~~1 &~~1 &-1& -1& -1& ~~1& ~~1& -1& -1& ~~1& ~~1 \\
\mathbf{z}_{2} &~~1 &-1 &~~1 &-1 &~~1 &~~1 &-1 &-1 &-1 &-1 &~~1 & ~~1\\
\end{block}
\end{blockarray}
\end{equation}
Crucially, it is important to convert our basis vectors in the FFF to the following $S$-matrix, from which we can calculate the partition function in the form discussed in Section \ref{sec:Partfunc}. We adopt the formalism developed
in ref. 
\cite{avalos2023d, Viktorthesis, FRtranslation} 
to convert from the FFF basis set phases to
the index set used in the expression for the partition function in eq. (\ref{eq:Z2})
\begin{equation}
S =  \text{  }  
\begin{blockarray}{ccccccccccccc}
& a & k & \rho & H_{1} & H_{2} & H_{3} & H_{4} & H_{5} & H_{6} & h_{1} & h_{2} & H\\
\begin{block}{c(cccccccccccc)}
\mathds{1} & 1 & 1 & 1 & 1 & 1 & 1 & 1 & 1 & 1 & 0 & 0 & 0 \\
\bm{\tilde{S}} & 1 & 0 & 0 & 0 & 0 & 0 & 0 & 0 & 0 & 0 & 0 & 0\\
\mathbf{e}_{1} & 0 & 0 & 0 & 1 & 0 & 0 & 0 & 0 & 0 & 0 & 0 & 0\\
\mathbf{e}_{2} & 0 & 0 & 0 & 0 & 1 & 0 & 0 & 0 & 0 & 0 & 0 & 0\\
\mathbf{e}_{3} & 0 & 0 & 0 & 0 & 0 & 1 & 0 & 0 & 0 & 0 & 0 & 0\\
\mathbf{e}_{4} & 0 & 0 & 0 & 0 & 0 & 0 & 1 & 0 & 0 & 0 & 0 & 0\\
\mathbf{e}_{5} & 0 & 0 & 0 & 0 & 0 & 0 & 0 & 1 & 0 & 0 & 0 & 0\\
\mathbf{e}_{6} & 0 & 0 & 0 & 0 & 0 & 0 & 0 & 0 & 1 & 0 & 0 & 0\\
\mathbf{b}_{1} & 1 & 1 & 0 & 0 & 0 & 0 & 0 & 0 & 0 & 1 & 0 & 0\\
\mathbf{b}_{2} & 1 & 1 & 0 & 0 & 0 & 0 & 0 & 0 & 0 & 0 & 1 & 0\\
\mathbf{z}_{1} & 0 & 0 & 1 & 0 & 0 & 0 & 0 & 0 & 0 & 0 & 0 & 1\\
\mathbf{z}_{2} & 0 & 0 & 0 & 0 & 0 & 0 & 0 & 0 & 0 & 0 & 0 & 1\\
\end{block}
\end{blockarray}
\end{equation}

The partition function for this model at the FFP, calculated to order $\mathcal{O}(2)$, is:
\begin{equation}
\begin{split}
    Z =& -200 + \frac{2}{\bar{q}} + 74560\bar{q} +\frac{448 \bar{q}^{\frac{1}{2}}}{q^{\frac{1}{2}}} +\frac{608\bar{q}^{\frac{5}{8}}}{q^{\frac{3}{8}}} +\frac{6912\bar{q}^{\frac{3}{4}}}{q^{\frac{1}{4}}} +\frac{33280\bar{q}^{\frac{7}{8}}}{q^{\frac{1}{8}}}-384 \bar{q}^{\frac{1}{8}}q^{\frac{1}{8}} \\
    & -2832\bar{q}^{\frac{1}{4}}q^{\frac{1}{4}} -9488\bar{q}^{\frac{3}{8}}q^{\frac{3}{8}}+\frac{184q^{\frac{1}{2}}}{\bar{q}^{\frac{1}{2}}}-16000\bar{q}^{\frac{1}{2}}q^{\frac{1}{2}}+\frac{432q^{\frac{5}{8}}}{\bar{q}^{\frac{3}{8}}}-16368\bar{q}^{\frac{5}{8}}q^{\frac{5}{8}} \\
    &+\frac{960q^{\frac{3}{4}}}{\bar{q}^{\frac{1}{4}}}+131200\bar{q}^{\frac{3}{4}}q^{\frac{3}{4}} +\frac{832q^{\frac{7}{8}}}{\bar{q}^{\frac{1}{8}}} +600768\bar{q}^{\frac{7}{8}}q^{\frac{7}{8}} -3264q \\
    &+ \frac{32q}{\bar{q}} + 1369472\bar{q}q  +\frac{8 q^{\frac{1}{4}}}{p^{\frac{3}{4}}}+\frac{16 q^{\frac{3}{8}}}{p^\frac{5}{8}}.
\end{split}
\end{equation}


\noindent
Integrating the partition function we find the model to have finite potential, $\Lambda$, at the FFP
\begin{equation}
    \Lambda = 0.08707 \mathcal{M}^{4}.
\end{equation}
The graphs are produced by spanning $T_{2}$ and $U_{2}$ in the region $\frac{1}{4} \leq T_{2}, U_{2} \leq 3$, and interpolating between these points. To this accuracy we find the minimum in potential sits at $(\frac{3}{4},\frac{3}{4})$, not at the FFP
\begin{equation}
    \Lambda_{\text{min}} = 0.08276 \mathcal{M}^{4}.
\end{equation}
Comparing these graphs to $S$--models in \cite{avalos2023d}, we can see that a feature they share is that the potential diverges as $T_{2}$ increase. In these models this is a result of the breaking of supersymmetry explicitly, not spontaneously. In this case the potentials of $\tilde{S}$-- and $S$--models share a similar distribution and become indistinguishable as $T_{2} \rightarrow \infty $ . 
\begin{figure}
    \centering
    \includegraphics[width=0.8\textwidth]{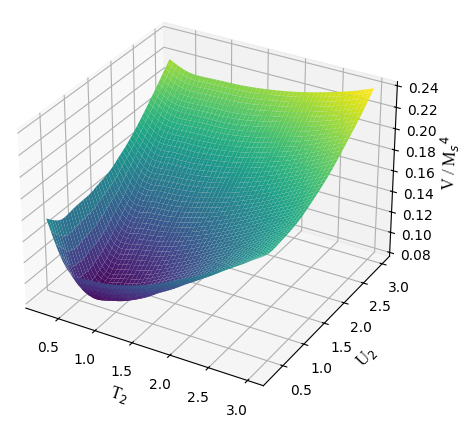}
    \caption{Graph to show an example of a stable $\tilde{S}$-Model, with a minimum in the potential, $V$, away from the FFP. The minimum in potential is located at $(\frac{3}{4},\frac{3}{4})$, with $V_{\text{min}} = \Lambda_{\text{min}} = 0.08276 \mathcal{M}^{4}$}
    \label{fig:T2U2}
\end{figure}
\begin{figure}
    \centering
    \includegraphics[width=0.8\textwidth]{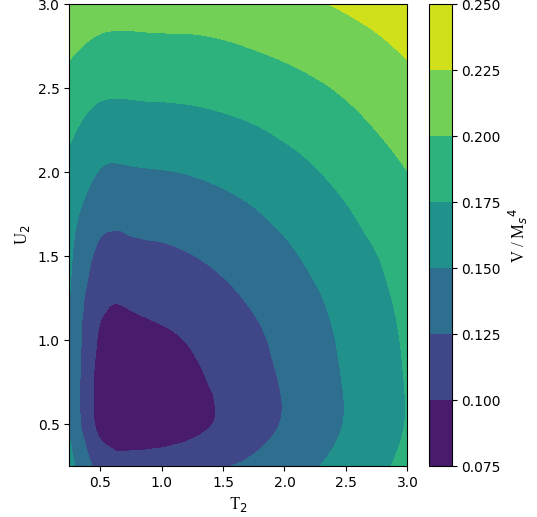}
    \caption{Contour plot of Figure \ref{fig:T2U2}}
    \label{fig:contour}
\end{figure}

\section{Discussion and Conclusion}
\label{sec:con}

In this paper we analysed the space of heterotic--string $\tilde{S}$--models in the Free Fermionic Formalism. These models are non--supersymmetric $SO(10)$ string models and so lose some of the appealing features of supersymmetric models. In general the spectrum will contain tachyonic states and thus be unstable at the FFP. Computer programs, written in C++, were developed and utilised to classify a random sample of $1.0 \times 10^{9}$ $\tilde{S}$--vacua against phenomenological conditions, in search of models compatible with SM data, and with similar features to previously studied $S$--models. Following a similar procedure to 
previous analysis \cite{faraggi2021classification, faraggi2020towards, florakis2016chiral}, these conditions are chosen to be: projection of all level-matched tachyons; accommodating $3$ spinorial generations of SM particles; production of $1$ vectorial state compatible with the Higgs doublet; TQMC survival; and having a finite, stable potential around the FFP. We also look for models that mimic supersymmetry with equal numbers of massless bosons and fermions.
In Section \ref{sec:Results} these results are outlined. Of the $ 10^{9}$ GGSO matrices which were sampled, only $0.5309\%$ were tachyon free. Of these, only $0.003072\%$ were compatible with the SM fermion generations and Higgs particle. $(8.4 \times 10^{-5})\%$ of our sample adhered to the TQMC conditions, with $4$ out of $10^{9}$ showing Bose-Fermi degeneracy at the massless level.

Crucially, in Section \ref{sec:Ex}, we adapted the notation and methodology of \cite{avalos2023d, avalos2023fayet} and applied it to non--supersymmetric 
$\tilde{S}$--models. We demonstrated that the $\tilde{S}$--models provide an abundance of non--supersymmetric vacua, which are both compatible with fertility conditions, and have finite one--loop potential at the FFP. Our sample suggests that $98\%$ of these vacua have positive Cosmological Constant, corresponding to a De Sitter space. This statistic is only based on a sample of $84$ models and so is likely to be inaccurate, however it does demonstrate that both positive and negative cosmological constants can be found. Fertile models can be constructed that sit at the minima of the potential at or around the FFP, when the geometric moduli $T^{(1)}_{2}$ and $U^{(1)}_{2}$ are varied. It is important to note that since the other geometric moduli were fixed at the FFP, we do not know the behaviour of the models when $T^{(1)}_{1}$ and $U^{(1)}_{1}$ are varied. Similarly, we can not speak to the behaviour of the model for values of $T^{(2,3)}$ and $U^{(2,3)}$ away from the FFP. 

In regard to the example model in Section \ref{sec:Ex} for which the minimum in potential sits away from the FFP, we cannot guarantee that the fertility conditions are satisfied at this point. We also can't say whether this is the global minimum across all moduli space. We can only conclude that the model remain tachyon free in the moduli space we have sampled, but further analysis of the spectrum away from the FFP would need to be undertaken.
Furthermore, we only calculate the potential at the one loop level. Another factor in the stability of the model and value of the potential is the existence of tadpole diagrams and uplifts \cite{avalos2023d, avalos2023fayet}. A thorough exploration of these effects on the model would be required before drawing further conclusions on stability of models, and their compatibility with cosmological data such as the cosmological constant. In this regard, this work simplifies the picture, but demonstrates that these models warrant further investigation.

A limiting factor to be addressed in this work is computational time. Randomly sampling such a large space is not the most efficient way to find fertile cores \cite{faraggi2021satisfiability}, but was necessary in this work to find a representative statistical sample. In exploring such a large space of vacua, we define the fertility conditions necessary to find consistent string vacua. 
This can act as a jumping off point should these models be analysed further, mitigating a portion of the computing time and producing an increased frequency of fertile models. For example, further work to break these models down to Pati-Salam \cite{faraggi2021classification} or Standard-like Models could be done to analyse the effect this has on the potential. 

 However, calculating the partition functions of the fertile models away from the FFP was the most inefficient aspect of this work. This is due to the large dimensionality and complexity of the calculation. Because of this, we are limited to only exploring the effect of $T^{(1)}_{2}$ and $U^{(1)}_{2}$ on the potential. An appealing solution would be to employ more advanced computational techniques such as machine learning, which are better suited for handling high-dimensional computation and categorisation problems, and are likely to increase efficiency. This should allow us to explore a greater proportion of the moduli space and perhaps lead to a greater understanding of the potential.

As there is no way to restore supersymmetry in $\tilde{S}$--models, it would seem the viability of these models relies on the absence of supersymmetry from Nature all the way up to the $SO(10)$ GUT level, and, whilst there is no evidence of supersymmetry at the time of writing, improvements to experiments at CERN and other proposed experiments 
\cite{faraggi2023string, narain2022future, european2020update} promise to probe higher energy levels. Should evidence of supersymmetry be discovered, this would raise questions on the relevance of non--supersymmetric 
${\tilde S}$--models to low scale physics. 
Nevertheless, we comment that exploration of the role of the 
tachyonic ten dimensional string vacua aims at developing an understanding
of the string scale, rather than their potential relevance to low scale 
physics. It should be viewed more in the context of the underlying 
theory that unifies the ten dimensional string vacua, or as it is traditionally
dubbed as $M$--theory. It seems plausible that within that context, the
tachyonic ten dimensional vacua may play a role.

\section*{Acknowledgements}
This research was supported in part by grant NSF PHY-2309135 to the Kavli Institute for Theoretical Physics (KITP). 
The work of LD is supported by STFC and LIV.INNO. The work of ARDA is supported in part by EPSRC grant EP/T517975/1.
AEF would like to thank the Kavli Institute for Theoretical Physics and 
the CERN theory division for hospitality.

\newpage

\printbibliography

\end{document}